\documentclass[conference, a4paper]{IEEEtran}
\IEEEoverridecommandlockouts
\usepackage{subfig}
\usepackage{cite}
\usepackage{amsmath,amssymb,amsfonts}
\usepackage{algorithmic}
\usepackage{graphicx}
\usepackage{textcomp}
\usepackage{eps topdf}
\usepackage{xcolor}
\def\BibTeX{{\rm B\kern-.05em{\sc i\kern-.025em b}\kern-.08em
    T\kern-.1667em\lower.7ex\hbox{E}\kern-.125emX}}
\usepackage{caption} 

\newcommand{\liu}[1]{{\color{black} #1}}
\newcommand{\sofie}[1]{{\color{black} #1}}

\begin{document}

\title{Sensing with OFDM Waveform at mmWave Band based on Micro-Doppler Analysis}

\long\def\symbolfootnote[#1]#2{\begingroup%
\def\thefootnote{\fnsymbol{footnote}}\footnote[#1]{#2}\endgroup}
\renewcommand{\thefootnote}{\fnsymbol{footnote}}
\author{\IEEEauthorblockN{Mingqing Liu\IEEEauthorrefmark{1}\IEEEauthorrefmark{2},~Fei Gao\IEEEauthorrefmark{3},
~Zhuangzhuang Cui\IEEEauthorrefmark{2},~Sofie Pollin\IEEEauthorrefmark{2}, and~Qingwen Liu\IEEEauthorrefmark{1}}\IEEEauthorblockA{\IEEEauthorrefmark{1}College of Electronic and Information Engineering, Tongji University, Shanghai, China\\
\IEEEauthorrefmark{2}WaveCoRE, Department of Electrical Engineering (ESAT), Katholieke Universiteit Leuven, Belgium\\
\IEEEauthorrefmark{3}Wireless Access Research Innovation, Nokia Bell Labs, Shanghai, China.
}
}


\maketitle

\begin{abstract}
Joint communication and sensing (JCAS) technology has been regarded as one of the innovations in the 6G network. \liu{With the channel modeling proposed by the 3rd Generation Partnership Project (3GPP) TR 38.901, this paper investigates the sensing capability using the millimeter-wave (mmWave) band with an orthogonal frequency
division multiplexing (OFDM) waveform. Based on micro-Doppler (MD) analysis, we} 
present two case studies, i.e., fan speed detection and human activity recognition, to demonstrate the target modeling with micro-motions, backscattering signal construction, and MD signature extraction using an OFDM waveform at 28~GHz. \liu{Simulated signatures demonstrate distinct fan rotation or human motion, and waveform} parameters that affect the MD signature extraction are analyzed. Simulation results \liu{draw the validity of the proposed modeling and simulation methods, which} 
also aim to facilitate the generation of data sets for various JCAS applications. 
\end{abstract}


\section{Introduction}
\label{sec:Introduction}
Demands for sensing on communication systems are expected for various applications in the sixth generation (6G) networks~\cite{tan2021integrated,cui2021integrating}. On the one hand, adopting communication signals for sensing can deal with the information security and deployment challenges of sensing based on cameras or sensors. On the other hand, sensing results can further assist in improving communication performance, i.e., coping with foreign object intrusion. Typical applications of sensing based on communication include high-precision target positioning and identification, imaging, environmental reconstruction, etc. However, the above applications are still challenging due to the poor resolution of radio frequency (RF) signals. Exploring channels with higher carrier frequency is increasingly prominent, where investigating the sensing capability is also appealing. One of the challenges in dynamic sensing lies in identifying moving targets, where micro-Doppler (MD) has attracted many interests. Taking MD signature as the sensing metric, this paper focuses on high-resolution sensing at the millimeter-wave (mmWave) band with orthogonal frequency division multiplexing (OFDM) waveform.

It has shown that MD signatures were utilized to improve the capability of target detection and identification for sensing on communication~\cite{chen2014radar}. From MD analysis, we can extract the micro-scale movement signatures of the target. For example, arms swing will produce Doppler frequency shifts in addition to frequency shifts introduced by a person's walking. Moreover, the MD signature can characterize micro-motion for a particular target. Hence it is exploited for identifying and classifying different targets, e.g., helicopter detection, human activity recognition, etc., in radar systems~\cite{chen2014radar,Bespoke,RCS1}. \liu{Towards the scope of JCAS, we primarily simulate MD signatures with an OFDM waveform. JCSA simulations generally impose the interaction between communication and sensing. Assuming a hybrid channel model where communication has no effects on sensing, we adopt tapped delay line (TDL) channel modeling method proposed by the 3rd Generation Partnership Project (3GPP) TR 38.901 for target sensing simulation~\cite{3gpp}.
Two case studies are presented: rigid rotating fan detection where MD modulation is explained intuitively and non-rigid human activity identification to specify practical application.
}



Existing works on OFDM waveform-based JCAS focus on waveform design to trade-off communication and sensing performances~\cite{frame,OFDMwave2,OFDMwave3}. A typical scenario for JCAS application faces on vehicle-to-vehicle (V2V) networks, where the location and velocity of vehicles are estimated~\cite{OFDMpilot,OFDMwave3}. Meanwhile, the Doppler and delay estimation methodologies with OFDM waveforms have been investigated in~\cite{OFDMpilot,OFDMwave3,OFDMwave4}. With 3GPP initializing the standardization of JCAS, range-Doppler estimation is also conducted using long-term evolution (LTE) and 5G new radio (NR) waveforms at the mmWave band~\cite{5Gnr}. \liu{However, conducting the MD analysis and sensing channel simulation towards JCAS still need more investigations.}

Besides, to realize MD-based sensing applications in human activity recognition, there are two solutions to first simulate human motions: i) a human walking model derived from extensive biomechanical experiments is used, where $12$ analytical expressions are established to control the motion trajectory of $17$ reference points, however facing difficulties in modeling other human motions besides human walking~\cite{walking,chen2014radar}; ii) the motion capture (Mocap) system is adopted to obtain the trajectory of realistic human motions, which is combined with the human skeleton model to derive human animations~\cite{Bespoke}. The open-access Mocap database, i.e., OpenPose from CMU~\cite{CMU}, Mocap Database HDM05~\cite{HDM05}, etc., allows us to simulate realistic human activities and backscattering signal generation.

Overall, this paper aims to present MD-based sensing with OFDM signals at mmWave band by modeling and simulating MD signatures of a rotating fan and \sofie{a} moving human. Our contributions 
\sofie{in} this manuscript are summarized as
\begin{itemize}
\item
We investigate 
OFDM waveform sensing at mmWave 
frequencies based on MD analysis. The backscattering signals from moving targets are constructed with a TDL-based channel model and the impact
of system parameters on MD signature extraction are analyzed.

\item
We introduce two case studies to 
analyze the MD signatures.
Simulated signatures from generated spectrograms depict different fan rotations and human motion, which also facilitates data set generation for JCAS system design and analysis.

\end{itemize}

The remainder of this paper is organized as follows. In Section II, we introduce range, Doppler, and MD estimation based on OFDM backscattering signal construction. In Sections III and IV, we conduct two case studies for simulating MD signatures of fan rotation and human movements. Simulation results in Section V demonstrate the potential of MD analysis-based JCAS. Finally, we conclude the paper in Section VI.

\section{OFDM Waveform and JACS Architecture}
\begin{figure}[t]
    \centering
\includegraphics[width=3.5in]{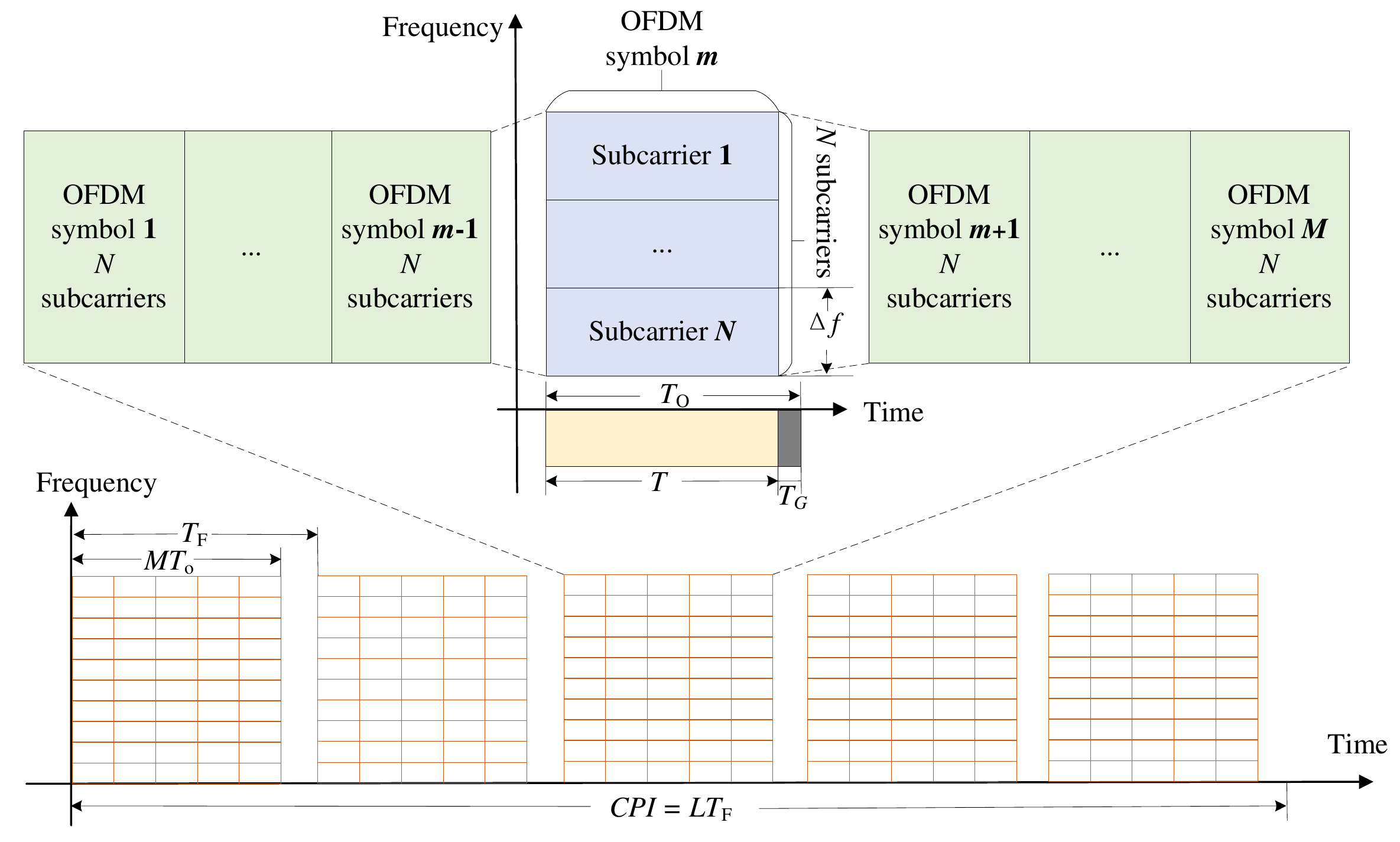}
    \caption{Schematic of an OFDM signal.}
    \label{f:frame}
\end{figure}

As a background, we first briefly introduce OFDM signal structure, transmission, backscattering signal construction, and range, Doppler, MD estimation. We also present assumptions for adopted OFDM-based JCAS architecture.

\subsection{OFDM Waveform Structure}
The bandwidth of the OFDM signal is divided into $N$ subcarriers, and one frame consists of $M$ symbols as shown in Fig.~\ref{f:frame}. The frequency spacing between subcarriers is denoted as $\Delta f$, and the symbol duration is $T = 1/\Delta f$.
Besides, a cyclic prefix is inserted before each OFDM symbol with a duration $T_G$ to avoid inter-symbol interference, where its typical values include $T_G = T/4$ or $T_G = T/8$. Thus, the total OFDM duration is expressed as $T_O=T+T_G$~\cite{phd}. Moreover, $T_F$ represents the package repetition interval (PRI) consisting of transmitting time for $M$ symbols and a preamble allocation time for Doppler estimation improvement~\cite{OFDMpilot}. $L$ packages are transmitted within a coherent processing interval (CPI). The time-domain data from the transmitter (Tx) is depicted as
\begin{equation}
\begin{aligned}
x(t)=&\frac{1}{N} \sum_{l=0}^{L-1}\sum_{m=0}^{M-1} \sum_{n =0}^{N-1} g\left(t-m T_{O}-lT_{\rm F}\right)\times\\
& S[n,m,l] \exp\left[{j 2 \pi n \Delta f\left(t-m T_{O}-lT_{\rm F}\right)}\right]
\end{aligned},
\label{e:txsignal}
\end{equation}
where $g(t)$ is the pulse shaping function and $S[n,m,l]$~is modulated symbols on a subcarrier, where indices $n,m,l$ indicate subcarrier, OFDM symbol, and package, respectively. Complex symbols $S[n,m,l]\in \mathcal{A}$ are from a modulation alphabet $\mathcal{A} \subset \mathbb{C}$, e.g., BPSK, QPSK, PSK, and QAM, etc. Here we intuitively explained Eq.~\eqref{e:txsignal} by a matrix representing one transmitted OFDM frame as~\cite{phd}
\begin{equation}
\mathbf{F}_{\mathrm{Tx}}=\left(\begin{array}{ccc}
S[0,0] & \cdots & S[0,M-1] \\
\vdots & \ddots & \vdots \\
S[N-1,1] & \cdots & S[N-1,M-1]
\end{array}\right),
\end{equation}
where each row and column represent a sub-carrier and an OFDM symbol of one transmitted frame, respectively.

\subsection{OFDM Backscattering Signal Construction}
There are two working modes for JCAS, i.e., mono-static and bi-static. For mono-static JCAS, the transmitted signal is used for both sensing and communication, and there is a sensing sniffer working as the receiver (Rx). Thus, $x(t)$ is reflected by the target user, and then received by the Rx, denoted as $y(t)$. Assuming $P$ scattering points are considered for a moving target, $y(t)$ is obtained similar to Eq.~\eqref{e:txsignal}:
\begin{equation}
\begin{aligned}
\left(\mathbf{F}_{\mathrm{Rx}}\right)_{n,m}=&\sum_{p=0}^{P-1}a_p\left(\mathbf{F}_{\mathrm{Tx}}\right)_{n,m} \cdot \exp{[j 2 \pi mT_O f_{D, p}]} \\&\cdot \exp{[-j 2 \pi \tau_{p} (n\Delta f + f_c) ]}+({\mathbf{Z}})_{n,m}
\end{aligned},
\end{equation}
where $a_{p}$ is the attenuation factor of reflection from $p$-th scattering point, depending on its distance $r_{p}$ and radar cross section (RCS) $\sigma_p$ relative to Tx. $f_{D, p}$ is the Doppler shift by a relative radial velocity $v_p$ of the $p$-th scattering point as
\begin{equation}
f_{D, p} = 2\frac{v_p}{c_0}f_c,
\end{equation}
where $c_0=3\times10^8$m/s is the speed of light and $f_C$ is the carrier frequency. Moreover, the signal is delayed by time $\tau_{p}$ taken for traveling to the $p$-th scattering point and back as
\begin{equation}
\tau_{p} = 2\frac{r_{p}}{c_0}.
\end{equation}
Finally, $({\mathbf{Z}})_{n,m}$ represents additive white Gaussian noise.

\subsection{Estimation Problem}
In Eq.~(3), the received signal contains the Doppler shift and time delay information we are interested in. Remove $\mathbf{F}_{\mathrm{Tx}}$ from $\mathbf{F}_{\mathrm{Rx}}$ by element-wise division, we finally obtain the matrix containing channel information as
\begin{equation}
\begin{aligned}
(\mathbf{F})_{n,m} = &\sum_{p=0}^{P-1}\frac{(\mathbf{F}_{\rm Rx})_{n,m}}{(\mathbf{F}_{\rm Tx})_{n,m}}=\sum_{p=0}^{P-1}a_p \exp\left[{j 2 \pi mT_O f_{D, p}}\right]\\ &\cdot \exp\left[{-j 2 \pi \tau_{p} (n\Delta f+f_c) }\right]+(\tilde{\mathbf{Z}})_{n,m}
\end{aligned},
\label{e:F}
\end{equation}
where $(\tilde{\mathbf{Z}})_{n,m}={({\mathbf{Z}})_{n,m}}/{\left(\mathbf{F}_{\mathrm{Tx}}\right)_{n,m}}$.
Conducting estimation algorithms onto $\mathbf{F}$, the velocity and range of the target are determined. Applying a fast Fourier transform (FFT) with a length of $H$ on each column over symbols and processing each row over subcarriers by inverse FFT (IFFT) with a length of $K$,  we can obtain the periodogram of the channel as
\begin{equation}
    (\mathbf{P})_{n, m}=\frac{1}{NM}\left|\sum_{k=0}^{K-1}\left(\sum_{h=0}^{H-1} (\mathbf{F})_{k, h} e^{-j 2 \pi \frac{h m}{H}}\right) e^{j 2 \pi \frac{kn}{K}}\right|^2,
\end{equation}
where $K$ and $H$ can be larger than $N$ and $M$ with zero-padding to improve the frequency estimation accuracy. Then, the range $\hat{r}$ and velocity $\hat{v}$ are estimated by finding the peak value of $\mathbf{P}$ and the indices $(\hat{n},\hat{m})$ correspond to the target state as $\hat{r} = {\hat{n}c_0}/({2\Delta f K})$ and $\hat{v} = {\hat{m}c_0}/({2 f_cT_OH})$, respectively. Note that the estimation is limited by the resolution of range and velocity as
\begin{equation}
   \Delta r = \frac{c_0}{2N\Delta f} \quad\text{and}\quad\Delta v = \frac{c_0}{2f_cMT_O},
\end{equation}
based on which we can choose appropriate OFDM parameters for various application scenarios.

More importantly, $\mathbf{F}$ contains motion information of a target with more than one scatter point. Thus, various Doppler shifts and time delays caused by various parts of the target can be estimated. For example, blades keep rotating while moving away from or to the transmitter, and arms/legs swing while a human is walking. To extract micro-Doppler (MD) signatures caused by these micro-motions from $\mathbf{F}$, time-frequency (TF) analysis should be applied to $\mathbf{P}$, among which short-time Fourier transform (STFT) is the most commonly used method. MD spectrum is obtained by processing each row over symbols of $\mathbf{F}$ with STFT as
\begin{equation}
    (\mathbf{D})_{n,m} = \frac{1}{N_D}\left|\sum_{i=0}^{N_D-1}\sum_{q=0}^{Q-1}(\mathbf{F})_{n,q+m\sigma}w(q)e^{-j2\pi\frac{iq}{Q}}\right|^2,
\end{equation}
where $i=0,1,...N_D-1$ is the frequency index of FFT with length $N_D$, $w(\cdot)$ is the window function with a length $Q$, and $\sigma$ is the time granularity of STFT. Phased modulation changing due to velocity changing over time is estimated across subsequent windows over symbols.
\liu{
\subsection{Assumptions for JCAS Architecture}
In this work, we assume sensing and communication signals are separated in time, and the sensing works in mono-static mode with a separate Tx/Rx antenna. Under this condition, sensing performance will not be affected by communication, where the channel modeling method in 3GPP TR38.901 can be adopted for sensing simulation. Moreover, the sensing is assumed with an OFDM waveform with a $LT_{\rm F}$ repetitive interval and the sensing target is in free space.
}
\begin{figure}[t]
    \centering
\includegraphics[width=3.3in]{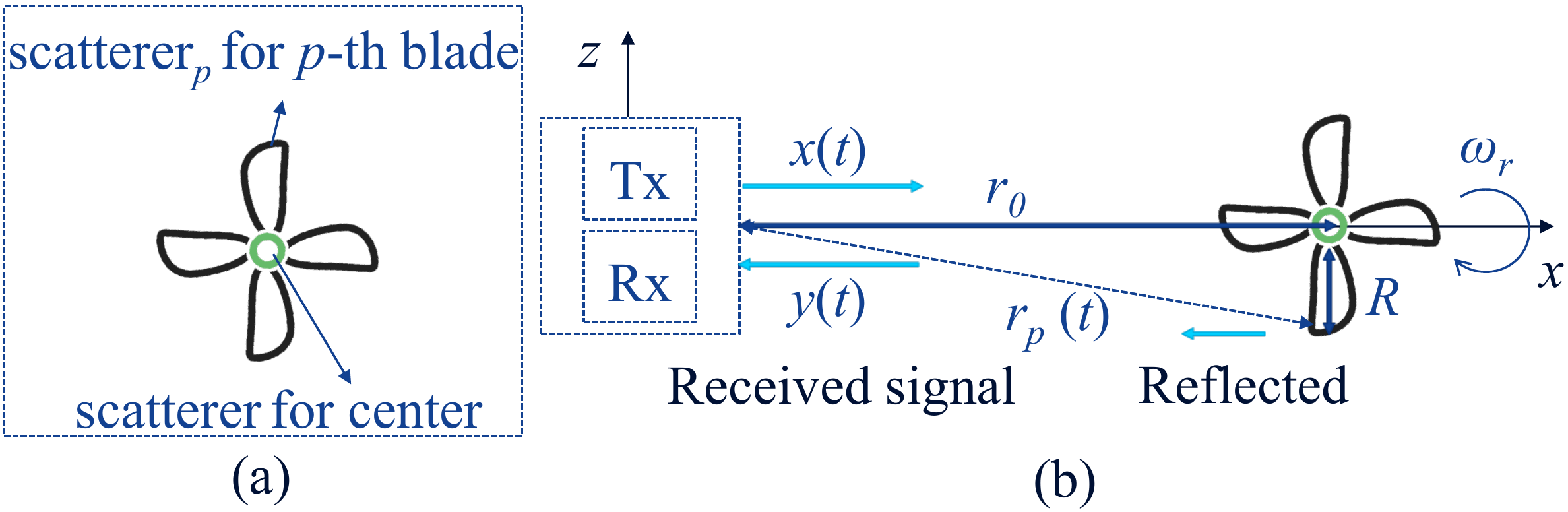}
    \caption{Model of fan and channel for sensing a rotating fan. (a) Illustration of scattering points on a fan, (b) Signalling process when a fan is a target.}
    \label{f:fan}
\end{figure}

\section{Case Study I: Fan Rotation Detection}
We first present the channel modeling  28~GHz 5G-mmWave band with OFDM waveform for fan rotation detection. The transmitted signal is in the form of Eq.~\eqref{e:txsignal}, and the 4-QAM modulation 
is adopted. Based on the TDL channel model, we emphasize modeling the rotating fan and constructing the backscattering signal under the fan's rotations.
\subsection{Fan Modeling with Micro-Motions}
Channel modeling for fan detection is depicted as follows. The first step is modeling the sensing target, that is, a fan as in Fig.~\ref{f:fan}(a). The fan is with four blades that have a length of $R$ and a central bearing, where the angular rotation velocity is $\omega_r$ (the blade rotating rate is equal to $3\omega_r/\pi$). The four blades share the same reflectivity, and the rotation center has higher reflectivity. Thus, five {scattering points} are set in a fan model. Besides, the {center of the} fan is placed at a distance of $r_0$ away from the mono-static {transceiver}. Then, the following assumptions of the channel in Fig.~\ref{f:fan}(b) are made: i) mono-static scenario is assumed with integrated Tx and Rx; ii) the line-of-sight (LoS) propagation is assumed for simplicity; and iii) the fan center is static relative to the transmitter.

Given the initial phase of each blade as $\varphi_p$, the distance between the transmitter and each blade changes with time due to the rotating of blades as
\begin{equation}
    r_p(t) = \sqrt{r_0^2+R^2+2r_0R\cos(\omega_{r} t+\varphi_p)},
\end{equation}
where $p=1,2,3,4$. Here the phase difference between every two adjacent blades is $\pi/2$.

\subsection{Backscattering Signal and MD signature Extraction}
To better align with the sampling process of OFDM signals, we sample the backscattering signal of a rotating fan at $t=mT_o+lT_{\rm F}$ and the matrix representing channel information is depicted as
\begin{equation}
\begin{aligned}
(\mathbf{F})_{n,m+lL} = &\sum_{p=0}^{P-1}a_p \exp\left[{j 2 \pi  f_{D, p}(mT_O+lT_{\rm F})}\right]\\ &\cdot \exp\left[{-j 2 \pi \tau_{p} (n\Delta f+f_c) }\right]+(\tilde{\mathbf{Z}})_{n,m}
\end{aligned},
\label{e:md}
\end{equation}
{where the velocity resolution and maximum detectable velocity are rewritten as} $\Delta v = c_0/(2f_cLT_{\rm F})$ and $v_{\rm max}=c_0/(4f_cT_{\rm F})$, respectively.

Conducting STFT on $\mathbf{F}$, we can obtain the spectrogram of the backscattering signal from a rotating fan. With the above modeling process, given the sensing system and fan parameters, we can {obtain} the spectrogram of the MD signature, so that we may identify the target as a fan from sinusoid-like MD frequency shifts due to blade rotation. It has shown that the modulation scheme will not affect the sensing performance as the modulated data is removed from the channel information. For a JCAS application, channel information is directly used for sensing, and the data flow is collected by the demodulator.
\begin{figure}[tbp]
	\centering
	\subfloat[] {\includegraphics[width=.23\textwidth]{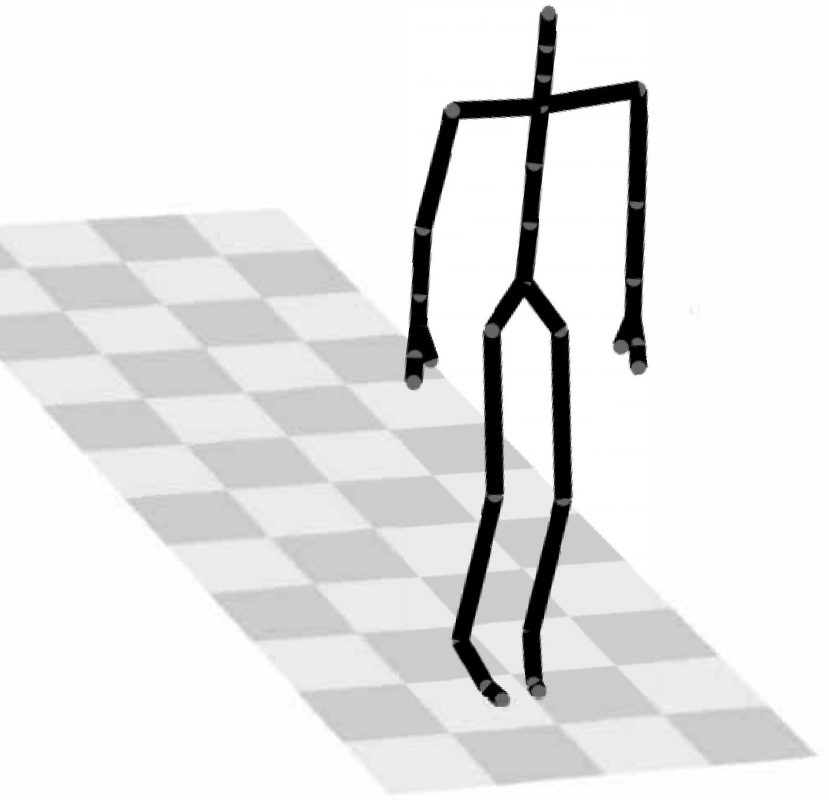}}
	\subfloat[] {\includegraphics[width=.19\textwidth]{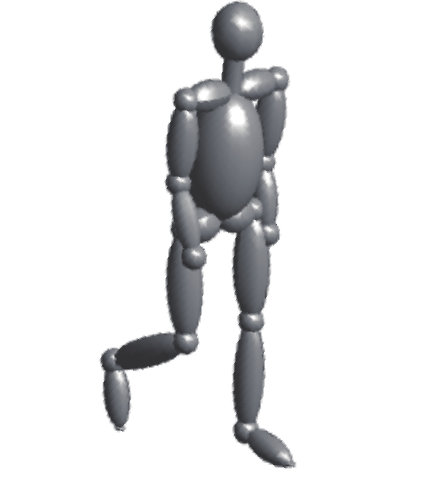}}
		\caption{The primitive shapes-based human motion model. (a) Joints on the human body. (b) Primitive shapes-based backscattering model.}
	\label{f:humanmodel}
\end{figure}
\section{Case Study II: Human Activity Recognition}
\label{sec:PositionModel}
Next, we simulate the MD signature for human motion with an OFDM waveform at mmWave frequencies.
In this section, we introduce the human motion model where the human skeleton modeling and actual trajectories combination are conducted. Moreover, we present the backscattering signal construction based on specifying RCS of different body parts with human motions and parameterization details for the OFDM waveform.
\subsection{Dynamic Human Motion Model}
Unlike modeling the rigid object, there are three challenges in building the dynamic human motion model as follows. Firstly, the human body has complex structures and rich reflection points. To address this issue, we adopted a schematic human body model as in Fig.~\ref{f:humanmodel}(a), which is built by a series of reference points indicating critical joints of the human body. Secondly, the movements and gestures of the human body are generally irregular, which 
makes it challenging to develop a corresponding mathematical model. In this paper, we adopt a Mocap Database HDM05 containing human trajectory data to simulate the kinematic and obtain the animation of human motions~\cite{HDM05}. Finally, human tissues on different parts have distinct reflection characteristics, and the equivalent area of each part seen by Tx, i.e., RCS, varies with time during human moving. We assume that the transmitted signal is completely reflected by each scattering reference point. Moreover, as in Fig.~\ref{f:humanmodel}(b), we use primitive shapes to model different parts of the human body and embody them into the human skeleton, e.g., using ellipsoids for the torso, arms, and legs and a sphere for the head, of which the RCS is easily emulated~\cite{chen2014radar}.

To build a dynamic human motion model, this paper combines the primitive shapes-based human model with $P=19$ shapes and data from HDM05 containing the motion information of the human skeleton. Assume the scattering centers lie approximately at the center of these primitive shapes, we can then derive the reflectivity $a_p(t)$ of each primitive (i.e., each body part) at any time instant $t$ is expressed by~\cite{RCS1,RCS2}
\begin{equation}
    \left\{\begin{aligned}
    &a_p(t) = \frac{\zeta(t) \sqrt{\sigma_p(t)}}{r_p^2(t)}\\
&\sigma_p(t)=\frac{\pi a_p^2 b_p^2 c_p^2}{\left(a_p^2 \Phi_p+b_p^2 \Psi_p+c_p^2 \cos ^2 \theta_p(t)\right)^2}\\
&\Phi := \sin ^2 \theta_p(t) \cos ^2 \phi_p(t)\\
&\Psi := \sin ^2 \theta_p(t) \sin ^2 \phi_p(t)
\end{aligned}\right.,
\end{equation}
where $\zeta(t)$ indicates propagation effects such as attenuation, $\sigma_p(t)$ and $r_p(t)$ are RCS of $p$-th primitive and distance away from radar. Specifically, RCS relies on the 
geometry of an ellipsoid with generalized parameters, where $a_p,b_p,c_p$ are radii along the $x,y,z$ axis, respectively. $\theta_p$ and $\phi_p$ are zenith and aspect angle, which depend on the time-varying relative positions between ellipsoid and radar due to human motions.

Thus, given a Tx/Rx location, distances $r_p$ between each body part and the Tx/Rx, as well as the RCS of each body part $\sigma_p$ during human moving are obtained with the established dynamic human motion model. Stated that for a bistatic configuration, computation for time-varying distances and RCSs is different, as illustrated in~\cite{RCS2,Bespoke}.
\subsection{Implementation with OFDM Waveform}
Combining the dynamic human model with motion information including trajectory and RCS of each body part, we can simulate the hybrid electromagnetic radar scattering from dynamic humans. However, further considerations need to be addressed during implementation. Human motion data from MOCAP HDM05 is sampled at $120$~Hz while the Doppler frequency shift generated by human motions is generally more significant, which lies in the $\pm1$~kHz Doppler range. Hence, we first interpolate HDM05 data to increase the sampling frequency to $2$ kHz, which should correspond to the JCAS Tx sampling frequency. Similar to fan simulation, we sample the signal at $t=mT_o+lT_{\rm F}$ and $LT_{\rm F}$ between every two packets determines the maximum measurable Doppler frequency; that is, $1/(LT_{\rm F})=2$~kHz.

Moreover, the channel information matrix is given as Eq.~\eqref{e:md}. Take a summation over subcarriers, $\mathbf{F}$ deduces to a vector on which STFT can conduct. Another option is selecting a vector from $\mathbf{F}$ over symbols via peak finding along the range, which we have adopted for fan rotation detection. For conducting STFT analysis, we adopt a Gaussian window with 600-point FFT and 95\% overlap.

\section{Numerical Results}
In this section, we present simulation results for generating MD signatures of fan rotation and various human motions. System and signal parametrization is conducted first to align with the OFDM waveform requirements. Then, we depict the spectrograms for MD signatures and illustrate features that are extracted from them. Sensing performance with different system or waveform parameters is also analyzed.
\begin{figure}[t]
    \centering
\includegraphics[width=3.3in]{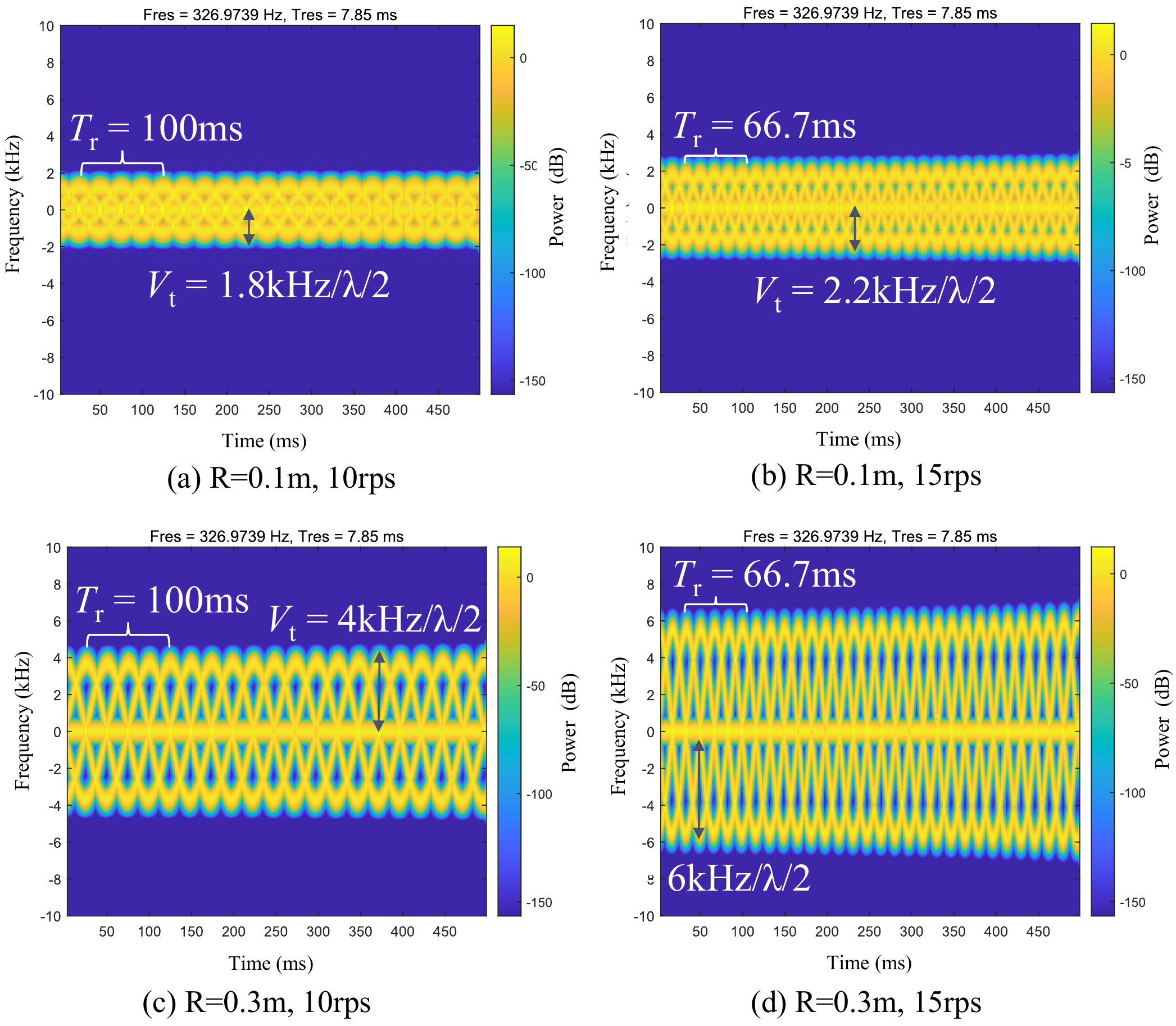}
    \caption{Spectrograms of Micro-Doppler signatures for a rotating fan at different rotational speeds and blade lengths.}
    \label{f:fanR1}
\end{figure}
\subsection{Simulation Parameters}
A mono-static configuration with carrier frequency $f_c = 28$~GHz is adopted in this paper. For fan detection, we consider a fan 
rotating at $10,15$ and $20$~rps. Thus, the packet repetition frequency is chosen as $20$~kHz. The blade length of a fan is set as $10$~cm and to achieve sufficient range resolution, we adopt $B=1.5$~GHz divided by $1024$ subcarriers. 
Moreover, the coordinate of Tx/Rx is $[0,0,0]$ and the initial coordinate of the fan is $[5,0,0]$. For indoor scenarios of human activity simulation, Tx/Rx is placed at $[0,3,0]$ while the human's original position is $[3,0.9,0]$. State that units of the above positions are all in meters.
The range resolution should be at least centimeter-level, where the minimum bandwidth $B$ should be $1.5$~GHz. The relative velocity of humans $v_{rel,{\rm max}}$ is generally less than $10$~m/s. Thus, the minimum sub-carrier spacing is at least one order of magnitude of $2v_{\rm rel}f_c/c_0=1.9$~kHz. Here we choose the number of subcarriers $N=1024$ and the number of OFDM symbols $M=16$. Considering the required packet repetition rate, the minimum $L$ is deduced. The parameterization is detailed in Table~\ref{t:fanPara}.

\subsection{MD Signatures for Fan Rotation}

We first plot spectrograms of MD signatures for a rotating fan at different rotational speeds with various blade lengths, as shown in Fig.~\ref{f:fanR1}. We can obtain the following information based on the spectrograms. i) As the image depicts four rotating sub-components and a central bearing causing Sinusoid-like micro-Doppler modulation, which is deemed as a typical feature of a fan. ii) The rotational speed is estimated from the period $T_{\rm r}$ of MD modulation due to one blade rotation. The rotational speed is $b_{\rm r} = (1/T_{\rm r})$rps, e.g., the rotational speed of $10$rps and $15$rps are easily estimated under given parameters in Figs.~\ref{f:fanR1}(a)-(d). iii) The blade length is derived from the maximum Doppler frequency. As a mono-static scenario assumed in this paper, the detected velocity of the blade tip is $V_{\rm t} = f_{D, \rm max}/(2\lambda)$. Then, the estimated blade length is $\hat{R}=V_{\rm t}/(2\pi b_{\rm r})$.

\begin{figure}[t]
    \centerline{
\includegraphics[width=3.3in]{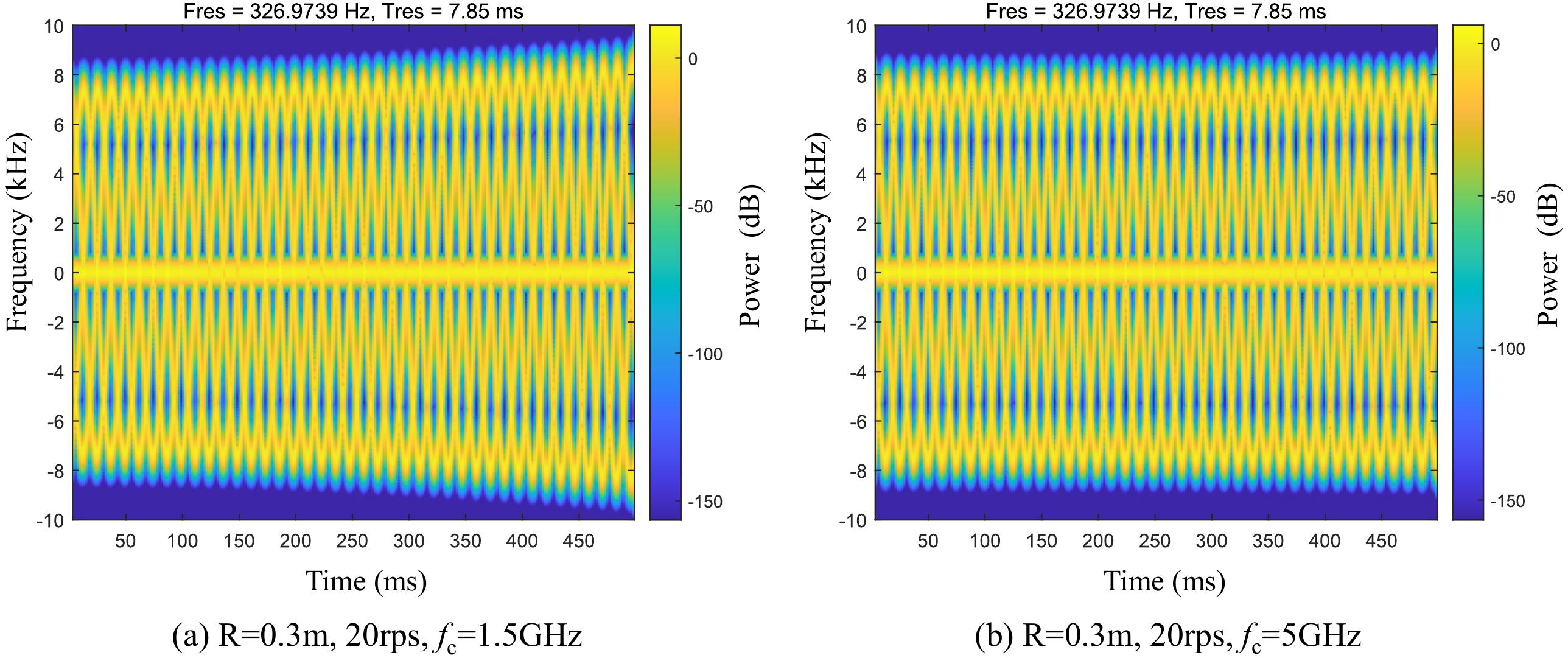}}
    \caption{Spectrograms of Micro-Doppler signatures for a rotating fan with different bandwidth.}
    \label{f:fanR2}
\end{figure}

Moreover, we find that smaller blade length with lower rotational speed causes difficulty in identifying each blade, as in Fig.~\ref{f:fanR1} (a). On the other hand, detecting a larger rotation speed requires a higher package repetition frequency. Besides, there exists signal distortion in Fig.~\ref{f:fanR1} (d), where the existing system parameter cannot guarantee the range and velocity resolution. Hence, we further analyze how the bandwidth impacts the MD signatures as in Fig.~\ref{f:fanR2}. For a rotating fan at a rotational speed of $20$rps and $0.3$m blade length as in Fig.~\ref{f:fanR2}(a), signal distortion is more obvious, and the MD signature may disappear if the bandwidth is smaller. However, we increase the bandwidth to $5$~GHz as in Fig.~\ref{f:fanR2} (b), and the MD signature recovers, which shows that high-precision sensing requires huge bandwidth, resulting in a trade-off design for JCAS.
\begin{table}[!htbp]
\caption{Parameter for JCAS Simulation}
\centering
\begin{tabular}{ccc}
\hline
\textbf{Parameter}&\textbf{Symbol}&\textbf{Value}\\
\hline
\text{Carrier frequency}&$f_c$&{\color{black}$28$GHz}\\
\text{Bandwidth}&$B$&{\color{black}$1.5$GHz}\\
\text{Number of subcarriers}&${N}$&{\color{black}$1024$}\\
\text{Number of OFDM symbols}&${M}$&{\color{black}$16$}\\
\text{Duration of a symbol}&${T}$&{\color{black}$0.68\mu$s}\\
\text{Duration of cyclic prefix}&${T_G}$&{\color{black}$0.17\mu$s}\\
\text{Total OFDM duration}&${T_o}$&{\color{black}$0.85$}\\
\text{Turn-off duration between frames}&${\rm T_{off}}$&{\color{black}$2.35\mu$s}\\
\text{Packet repetition interval}&${P_{\rm F}}$&{\color{black}$15.59\mu$s}\\
\text{Packet repetition frequency}&${\rm PRI}$&{\color{black}$2$kHz}\\
\text{Coherent processing interval}&${\rm CPI}$ & {\color{black}$0.5$ms}\\
\text{Number of packages}&$L$&{\color{black}$5000$}\\
\text{Maximum Doppler}&$f_{D,\rm max}$ & {$\pm1$kHz}\\
\text{Velocity resolution}&$\Delta v$ & {\color{black}$0.1$m/s} \\
\text{Range resolution}&$\Delta r$ & {\color{black}$6.7$cm} \\
\hline
\label{t:fanPara}
\end{tabular}
\end{table}

\subsection{MD Signatures for Human Motion}
\begin{figure}[htbp]
	\centering
	\subfloat[]{\includegraphics[width=.24\textwidth]{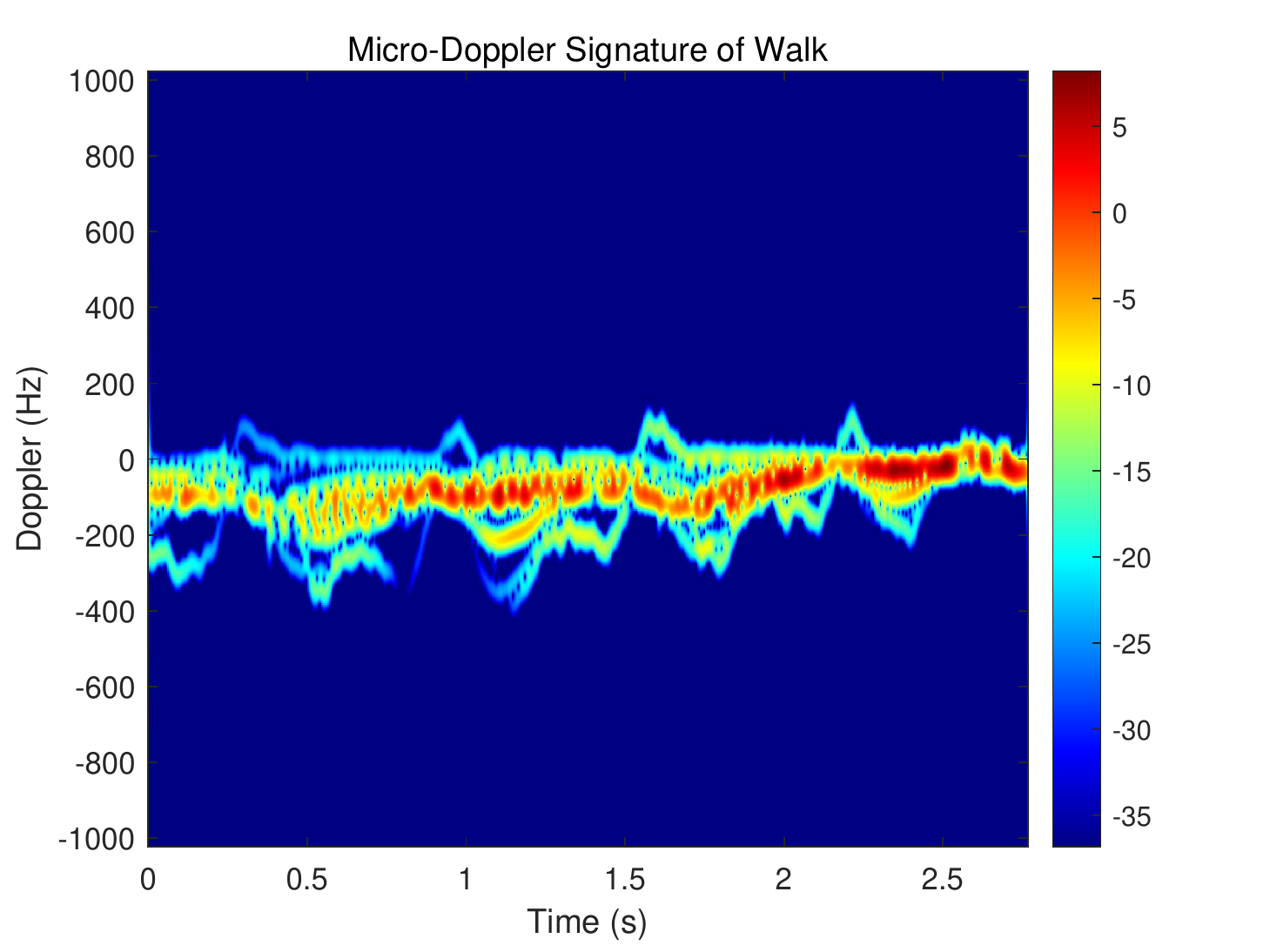}}
	\subfloat[]{\includegraphics[width=.24\textwidth]{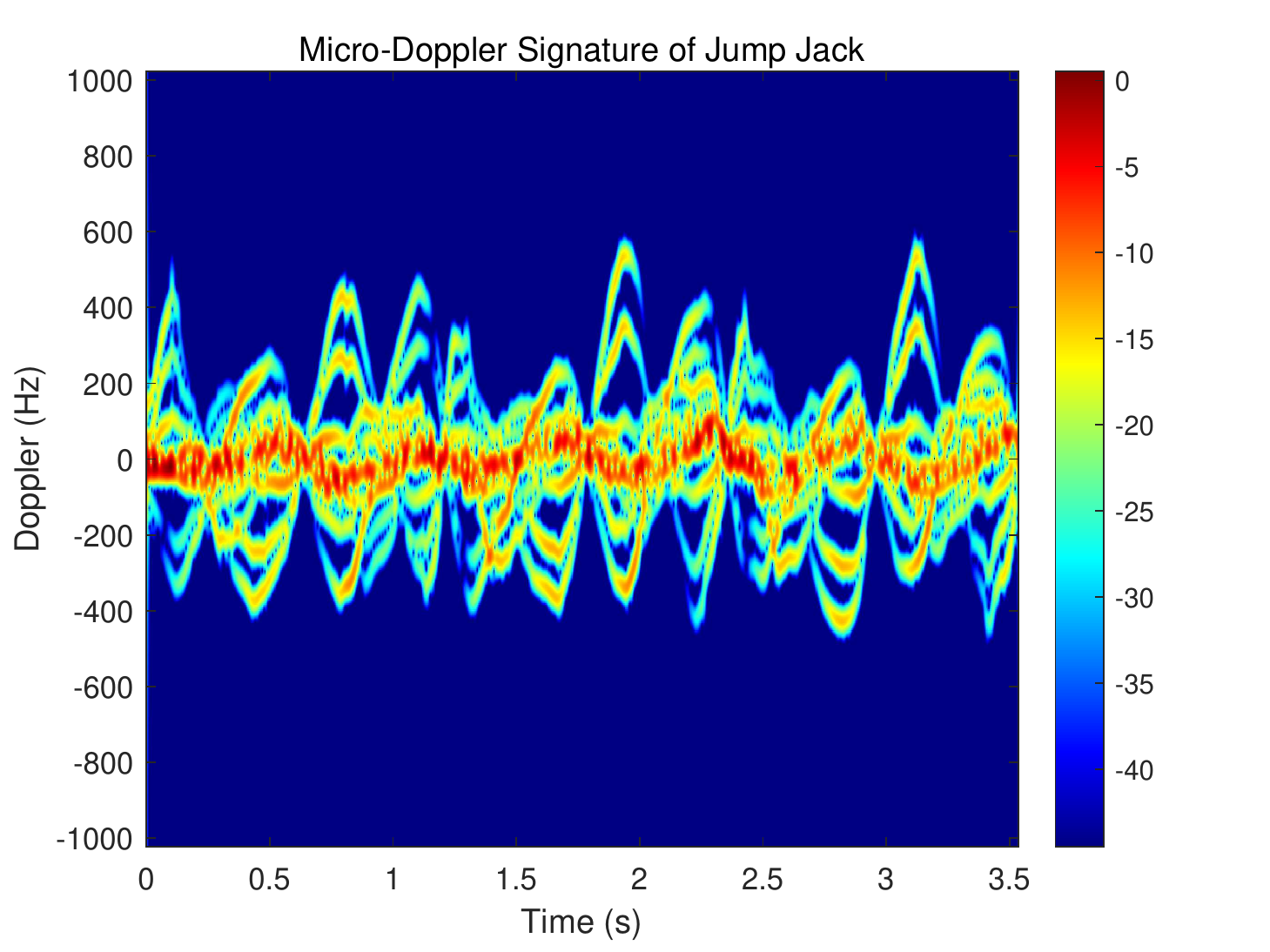}}\\
 	\subfloat[]{\includegraphics[width=.24\textwidth]{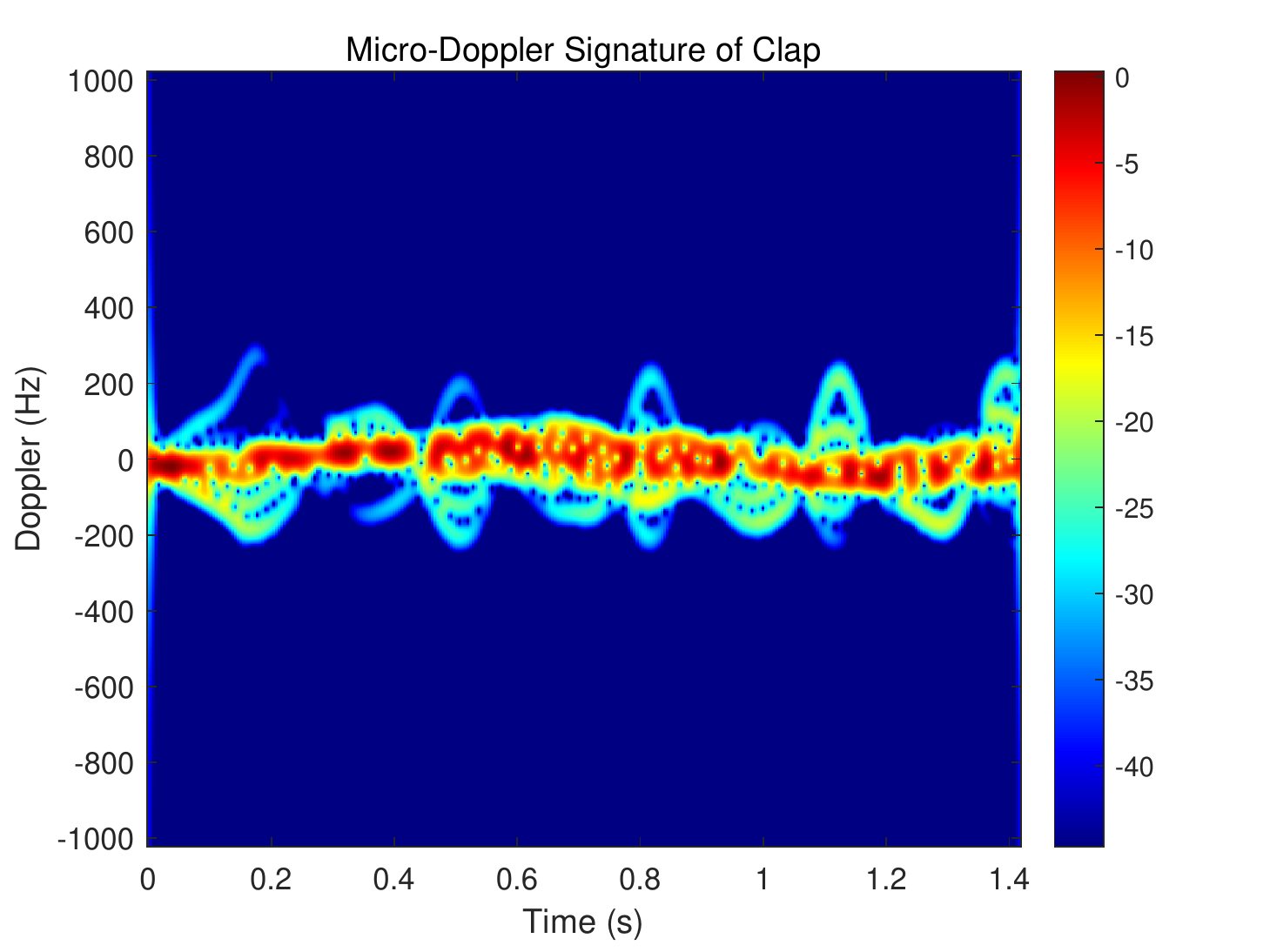}}
  \subfloat[]
   	{\includegraphics[width=.24\textwidth]{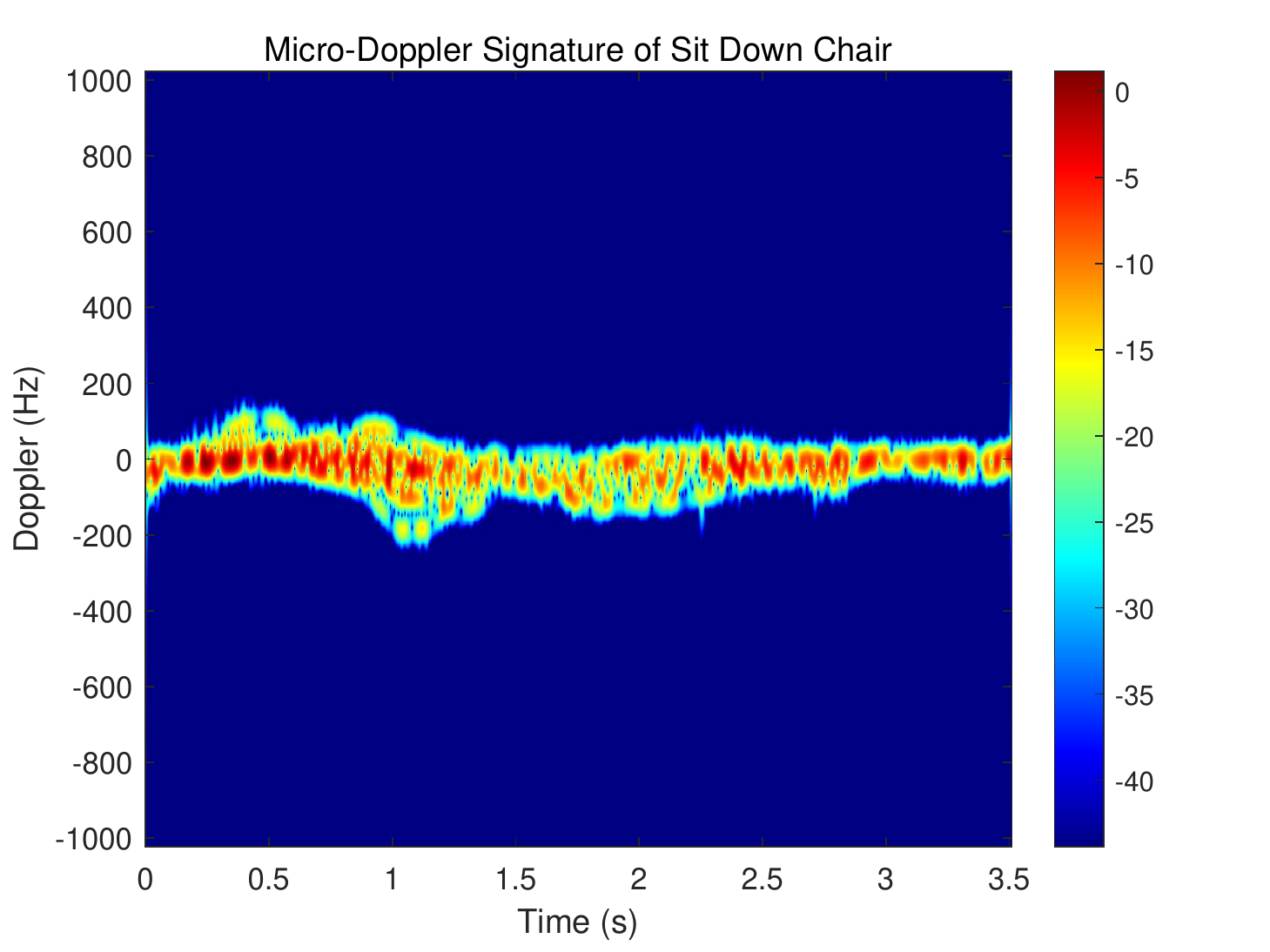}} \\
       	\subfloat []{\includegraphics[width=.24\textwidth]{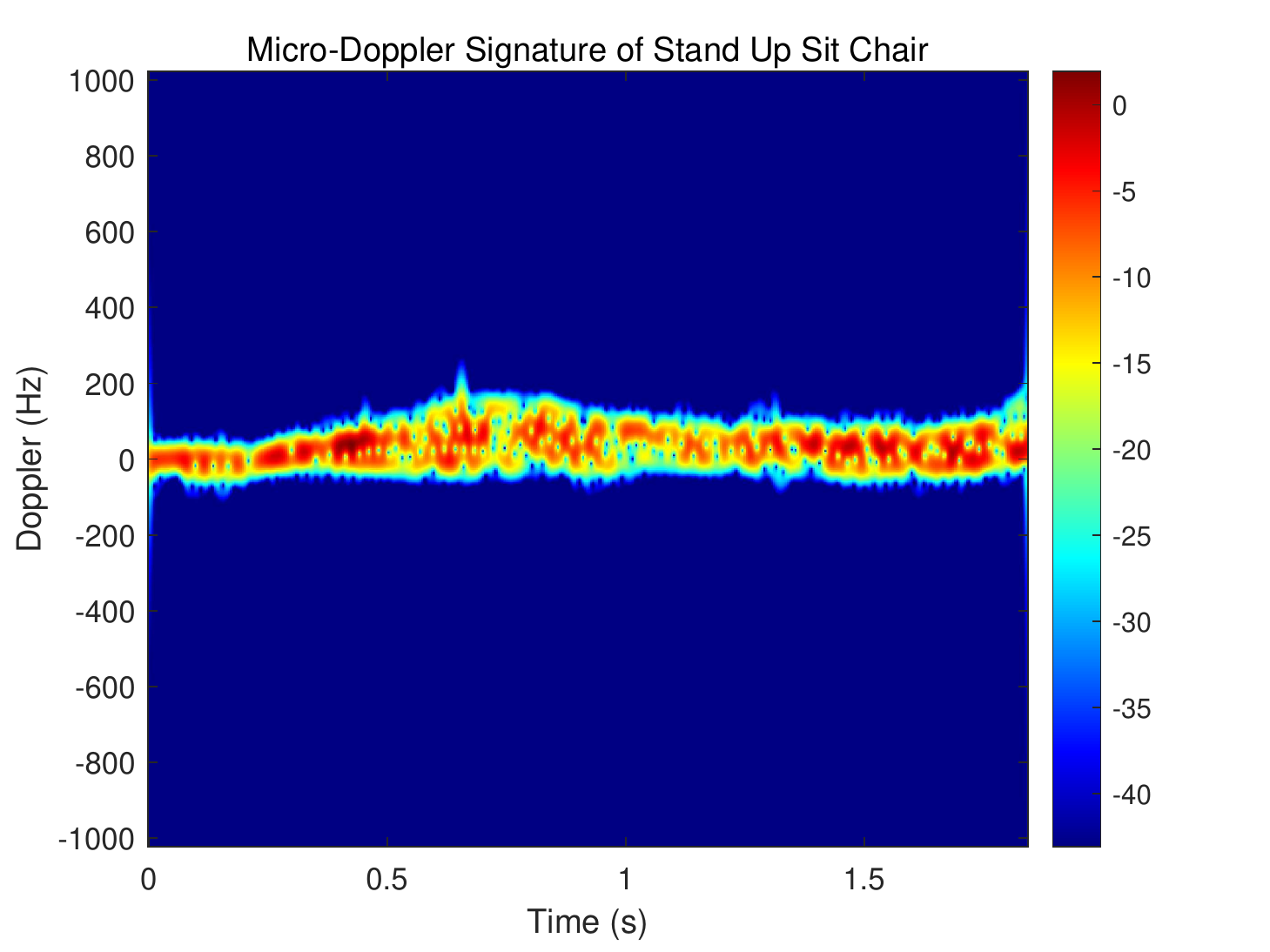}}
               	\subfloat []{\includegraphics[width=.24\textwidth]{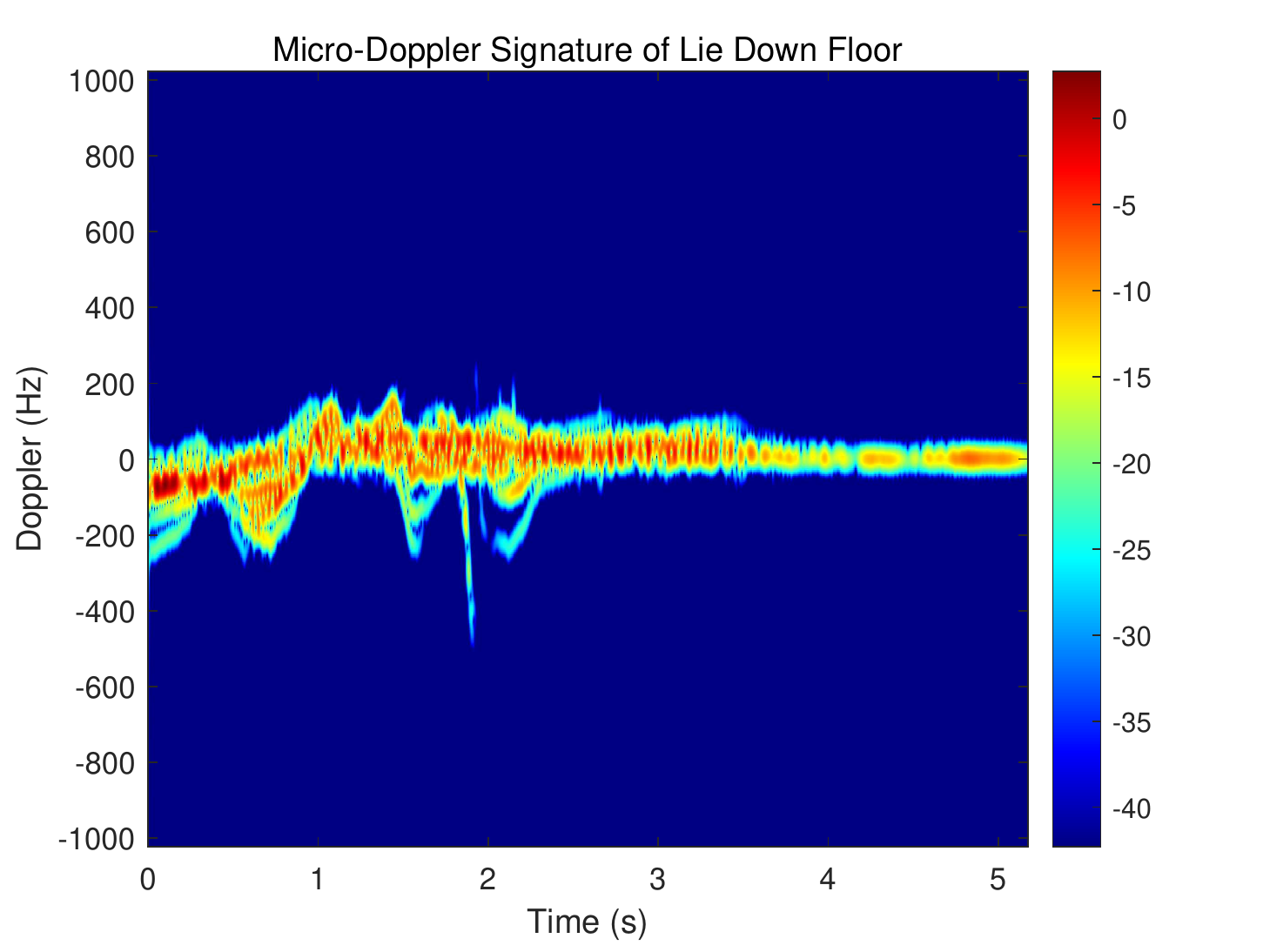}}
		\caption{Spectrograms of Micro-Doppler signatures with
the mono-static configuration for different motions: (a) Walking, (b) Jump jack, (c) Clap, (d) Sit down chair, (e) Stand up, (f) Lie down the floor.}
	\label{f:md}
\end{figure}

Spectrograms of MD signatures for different human activities using our OFDM waveform-based simulation are depicted in Fig.~\ref{f:md}. Due to the micro-motions of different body parts, MD features can indicate the particular human’s motions, which we can utilize for motion or activity identification and classification. Stated that Fig.~\ref{f:md} only shows some exemplary human motions by one actor from the database. However, the database contains more samples, and spectrogram segmentation can be conducted to obtain more data for algorithms such as machine learning~\cite{Bespoke}. Moreover, the signal processing for obtaining the above MD signatures is quite easily implemented. Thus, for home-caring scenarios, monitoring of a lie down floor activity as shown in Fig.~\ref{f:md}(f) is necessary.

\section{Conclusions}
\label{sec:Conclusion}
This paper presents the MD signature simulation with an OFDM waveform at mmWave bands. We illustrate the MD analysis-based backscattering signal construction and signal processing for MD signature extraction by conducting two case studies. Simulations have been established, where the relationships between MD signatures and specific target features to be estimated, as well as bandwidth requirements of sensing capability, are demonstrated. Our simulation results confirm that the MD reconstruction from standard OFDM signals can result in distinct signatures for human motion or fan rotation. From our results, we conclude that our signal reconstruction approach is valid and simulated signatures can be used for further MD-analysis-based JCAS system design. Several interesting topics are worthy of investigation in the
future: i) performance tradeoff for JCAS based on MD signature; ii) system design for improving JCAS performance; iii) sensing performance analysis under non-line-of-sight propagation.



\section*{Acknowledgements}

This work is partly supported by the 6G-BRICKS (Building Reusable testbed Infrastructures for validating Cloud-to-device breaKthrough technologieS) project under the European Union’s Horizon Europe research and innovation programme with Grant Agreement no. 101096954, and partly by KU Leuven Postdoctoral Mandate (PDM) under project no. 3E220691.

\bibliographystyle{IEEEtran}
\bibliography{Reference}

\end{document}